# Rate equation analysis and non-Hermiticity in coupled semiconductor laser arrays

Zihe Gao,[1] Matthew T. Johnson,[2] and Kent D. Choquette[1,a]

[1]*Department of Electrical and Computer Engineering, University of Illinois, Urbana, Illinois, 61801, USA*

[2]*United States Air Force Academy, Colorado Springs, Colorado, 80840, USA*

Optically-coupled semiconductor laser arrays are described by coupled rate equations. The coupled mode equations and carrier densities are included in the analysis, which inherently incorporate the carrier-induced nonlinearities including spatial hole burning and amplitude-phase coupling. We solve the steady-state coupled rate equations and consider the cavity frequency detuning and the individual laser pump rates as the experimentally controlled variables. We show that the carrier-induced nonlinearities play a critical role in the mode control, and we identify gain contrast induced by cavity frequency detuning as a unique mechanism for mode control. Photon-mediated energy transfer between cavities is also discussed. Parity-time symmetry and exceptional points in this system are studied. Unbroken parity-time symmetry can be achieved by judiciously combining cavity detuning and unequal pump rates, while broken symmetry lies on the boundary of the optical locking region. Exceptional points are identified at the intersection between broken and unbroken parity-time symmetry.

## I. INTRODUCTION

Coherent optically-coupled semiconductor laser arrays have been studied experimentally and theoretically for more than four decades [1-8]. Coupled mode theory has been successful in describing the optical coupled modes and the mutual coherence in coupled laser arrays [9-13]. Coupled rate equations (CREs) combine coupled mode theory with semiconductor laser rate equations, and have been used for the study of temporal dynamics of optically-coupled semiconductor laser arrays [6, 14]. In addition to capturing temporal dynamics, the CRE analysis also incorporates carrier-induced nonlinearities [15, 16], for example the gain saturation and the amplitude-phase coupling (i.e. nonzero linewidth enhancement factor resulting from carrier-induced frequency shift) [6, 17]. In this paper, we will show that these nonlinearities are critical not only for temporal dynamics, but also for the control of the steady-state coupled modes. By solving the steady-state coupled rate equations (SSCREs), we investigate the control mechanism for the relative intensity distribution from the elements of the array into the array supermode and the relative phase between cavities. We show that the control mechanism is governed by the carrier-induced nonlinearities, and the inclusion of carrier density is required in the analysis.

The phase tuning mechanism in optically-coupled semiconductor lasers has been a question of longstanding interest [18-22]. In the case of a real-valued coupling coefficient (for example arising from passive evanescent coupling), coupled mode theory shows that the gain contrast between lasers causes phase tuning, while the frequency detuning between cavities results in asymmetrical intensity distribution [18, 22, 23]. On the other hand, previous CRE analysis concludes the opposite in that frequency detuning is found to cause phase variation, but has negligible effect on intensity distribution [14, 21]. The latter is also in agreement with experimental observations suggesting that the frequency detuning causes the relative phase tuning [24]. In this paper, by carefully examining the cavity detuning and the total frequency detuning, we show that the two perspectives in fact do not contradict. We define the cavity detuning $\Delta\Omega$ to be the frequency detuning that excludes the contribution from amplitude-phase coupling, and we define the total frequency detuning $\Delta\omega$ to be the detuning that includes the amplitude-phase coupling, which is dependent on the actual carrier density distribution in the array.

In this paper we also apply our CRE analysis to parity-time (PT) symmetry and exceptional points in this optically-coupled non-Hermitian system. Comparing with previous PT symmetry analysis where gain saturation and frequency perturbation have been considered [25-29], we show that the amplitude-phase coupling is another nonlinearity that can play a critical role in optically-coupled semiconductor lasers in the weak coupling regime. As an addition to the well-known pump-induced PT symmetry breaking and exceptional points [25, 30], we demonstrate PT symmetry breaking induced exclusively by cavity detuning, as well as exceptional points induced by judiciously combining unequal pumping and cavity detuning.

## II. COUPLED RATE EQUATIONS

Assuming a real-valued coupling coefficient $\kappa$, the CREs can be written in terms of dimensionless variables [14]:

$$\frac{dY_A}{dt} = \frac{1}{2\tau_p}(M_A - 1)Y_A - Y_B \kappa \sin\phi \qquad (1)$$

$$\frac{dY_B}{dt} = \frac{1}{2\tau_p}(M_B - 1)Y_B + Y_A \kappa \sin\phi \qquad (2)$$

$$\frac{d\phi}{dt} = \frac{\alpha_H}{2\tau_p}(M_A - M_B) - \Delta\Omega + \kappa \cos\phi \left(\frac{Y_A}{Y_B} - \frac{Y_B}{Y_A}\right) \qquad (3)$$

$$\frac{dM_{A,B}}{dt} = \frac{1}{\tau_N}\left[Q_{A,B} - M_{A,B}(1 + Y_{A,B}^2)\right] \qquad (4)$$

Equations (1) - (3) represent coupled mode theory, where $Y_{A,B}$ are the normalized field magnitudes, and $M_{A,B}$ are the normalized carrier densities in cavity A and B, respectively. Furthermore $\phi \equiv \phi_B - \phi_A$ is the phase difference between the fields in cavity B and A, $\Delta\Omega \equiv \Omega_B - \Omega_A$ is the cavity detuning between lasers B and A, $\alpha_H$ is the linewidth enhancement factor, and $\tau_p$ is the photon lifetime. We have assumed that the real-valued coupling coefficients are symmetric $\kappa_{AB} = \kappa_{BA} = \kappa$, corresponding to two identical lasers that are passively coupled. Equations (1)-(3) are

---

a) Author to whom correspondence should be addressed. Choquett@illinois.edu

equivalent to the more familiar form of coupled mode theory written in terms of complex-valued field amplitudes [9, 10, 12], except that we have dropped the global phase and have kept only the relative phase $\phi$, as the global phase can be arbitrarily defined. Note that Equation (4) are the carrier density rate equations, where $Q_{A,B}$ are the normalized pumping rates in A and B, and $\tau_N$ is the carrier lifetime.

We have followed Ref. 14 and defined the normalized carrier densities, pump rates, and field magnitudes as:

$$M_{A,B} \equiv 1 + \frac{c}{n_g}\Gamma a_{diff}\tau_p(N_{A,B} - N_{th}) \quad (5)$$

$$Q_{A,B} \equiv C_Q\left(\frac{I}{I_{th}} - 1\right) + \frac{I}{I_{th}} \quad (6)$$

$$Y_{A,B} \equiv \sqrt{\frac{c\, a_{diff}\tau_N}{n_g}}|E_{A,B}| \quad (7)$$

where $N_{A,B}$ are the carrier densities, $N_{th}$ the threshold carrier densities, $P_{A,B}$ the pump rates, $n_g$ the group index, $\Gamma$ the confinement factor, and $a_{\text{diff}}$ is the differential gain. $C_Q$ is the constant relating injected currents to normalized pump parameters, defined as $C_Q \equiv \frac{a_{\text{diff}}N_{tr}}{g_{th}}$, where $N_{tr}$ is the transparency carrier density. The threshold gain $g_{th}$ is related to photon lifetime by $\frac{c}{n_g}\Gamma g_{th} = \frac{1}{\tau_p}$. The normalized parameters at transparency and threshold conditions are simply: $M_{A,Btr} = 0$, $M_{A,Bth} = 1$, $Q_{A,Btr} = 0$, $Q_{A,Bth} = 1$, where the subscript $tr$ denotes transparency and $th$ denotes threshold.

Unlike the case of a single laser, where its steady-state carrier density above threshold is pinned at the threshold value $N_{th}$, the carrier densities in each coupled laser in the array can be different from $N_{th}$. For example, one laser may have its carrier density higher than $N_{th}$, while the other laser has lower than $N_{th}$. Because of the amplitude-phase coupling via the linewidth enhancement factor in semiconductor lasers (caused by the dependence of the refractive index on the carrier density), the actual cavity resonance frequency $\omega_{A,B}$ will depend on the carrier density (and hence the cavity gain):

$$\omega_{A,B} = \Omega_{A,B} + \frac{\alpha_H(M_{A,B} - 1)}{2\tau_p} = \Omega_{A,B} + \alpha_H\gamma_{A,B} \quad (8)$$

where $\Omega_{A,B}$ are defined as the cavity resonance frequency when the carrier densities are pinned at the threshold level. The terms $\gamma_{A,B} \equiv (M_{A,B} - 1)/2\tau_p$ are the net gain (or loss if negative) in the cavity A or B. All frequency tuning mechanisms (e.g. thermal tuning of the cavity index) are included in $\Omega_{A,B}$, except for the amplitude-phase coupling. Amplitude-phase coupling is separately treated in Equation (8) by the term $\alpha_H\gamma_{A,B}$. When the lasers are not coupled, $\Omega_{A,B}$ and $\omega_{A,B}$ are always the same because of gain (carrier density) pinning. But when the lasers are optically coupled, $\Omega_{A,B}$ and $\omega_{A,B}$ are different. When the lasers are optically coupled, we can still vary $\Omega_{A,B}$ with thermal index tuning for example, but $\omega_{A,B}$ will differ from $\Omega_{A,B}$, because the laser array has the freedom of redistributing its carrier densities through photon-mediated energy transfer between elements as discussed later.

Setting the time derivatives in Equations (1) – (4) to zero, we get SSCREs, which are five algebraic equations with five real-valued unknowns ($Y_{A,B}, \phi, M_{A,B}$). We consider the pump parameters $Q_{A,B}$ and the cavity detuning $\Delta\Omega$ to be experimentally controlled and measurable. The terms $Q_{A,B}$ are directly related to the injected current through Equation (6), and $\Delta\Omega$ can be estimated by extrapolating the frequency shift in the uncoupled region [21, 31]. Approximate analytical solutions to the SSCREs can be found for equal pumping ($Q_A = Q_B$) assuming very weak coupling ($\kappa \ll 1/\tau_p$), as recently reported in Ref. 14. In general, there is no analytical solution for the SSCREs, particularly for coupled lasers with coupling coefficient comparable to the cavity loss rate, which is the case for coupled vertical cavity surface emitting laser (VCSEL) arrays [21, 32]. Numerical root search is used when analytical solution is not available. In addition to solving for the coupled optical modes, we also examine the tuning mechanism by calculating the gain contrast $\Delta\gamma$ and the total frequency detuning $\Delta\omega$ between cavities. They are related to the carrier density distribution through the following equations:

$$\Delta\gamma \equiv \gamma_B - \gamma_A = \frac{M_B - M_A}{2\tau_p} \quad (9)$$

$$\Delta\omega \equiv \omega_B - \omega_A = \Delta\Omega + \frac{\alpha_H}{2\tau_p}(M_B - M_A) \quad (10)$$

The device parameters used in this paper are $\alpha_H = 4$, $\tau_p = 2ps$, $a_{\text{diff}} = 5 \times 10^{-16} cm^2$, $N_{tr} = 2 \times 10^{18} cm^{-3}$, $n_g = 4$, $\Gamma = 0.04$, $C_Q = 0.6$, which are typical values for VCSELs [33]. Two values for the coupling coefficient are considered, which we denote as Array 1 and Array 2 in the following analysis. Array 1 has $\kappa = 1 \times 10^9 rad/s = \frac{0.002}{\tau_p} \ll \frac{1}{\tau_p}$, while Array 2 has $\kappa = 30 \times 10^9 rad/s = \frac{0.06}{\tau_p}$. Array 1 is in the limit of very weak coupling, while the coupling in Array 2 is stronger, being an experimentally estimated value for coupled VCSEL arrays under consideration [21]. Note that both cases are in the weak coupling regime, meaning the photons leak out of the system faster than interacting with the other cavity ($\kappa < 1/\tau_p$). This is in contrast to other optically-coupled laser systems where $\kappa > 1/\tau_p$ [30, 34].

## III. VERY WEAKLY COUPLED ARRAY UNDER EQUAL PUMPING

We first consider two semiconductor lasers that are very weakly coupled (Array 1) and equally pumped ($Q_A = Q_B$). The approximate analytical solution of the SSCREs, accurate to the first order of small $\tau_p\kappa$, was reported in Ref. 14, and is repeated here:

$$sin\phi \cong \frac{\Delta\Omega}{2\alpha_H\kappa} \quad (11)$$

$$M_A \cong 1 + 2\tau_p\kappa sin\phi \quad (12)$$

$$M_B \cong 1 - 2\tau_p\kappa sin\phi \quad (13)$$

$$Y_A^2 \cong Q(1 - 2\tau_p\kappa sin\phi) - 1 \quad (14)$$

$$Y_B^2 \cong Q(1 + 2\tau_p\kappa sin\phi) - 1 \quad (15)$$

From Equations (11)-(13), we know the carrier density distribution of the array as a function of cavity detuning $\Delta\Omega$. Using Equations (9) and (10), we can calculate the gain contrast and the total frequency detuning between cavities:

$$\Delta\gamma \cong -\frac{\Delta\Omega}{\alpha_H} \quad (16)$$

$$\Delta\omega \cong 0 \quad (17)$$

Equations (16)- (17) demonstrate that the cavity detuning $\Delta\Omega$ induces a proportional gain contrast, but the total frequency detuning is negligible. Equations (16) and (17) elucidate why the two explanations for the origin of phase tuning do not contradict. It is true that $\Delta\gamma$ controls the phase tuning, and $\Delta\omega$ controls the intensity distribution, as derived from the coupled mode theory [22]. But from the CRE perspective, we see that the cavity detuning $\Delta\Omega$ induces a proportional gain contrast $\Delta\gamma$, and hence it influences the beam steering through the induced gain contrast. On the other hand, the total frequency detuning $\Delta\omega$ is almost completely balanced by the frequency shift due to the asymmetric carrier distribution, and hence $\Delta\Omega$ has little effect on the intensity distribution.

When calculating eigenmodes of the laser array using coupled mode theory, the input is gain contrast $\Delta\gamma$ and total frequency detuning $\Delta\omega$, neither of which can be easily measured experimentally. Hence an advantage of CRE analysis is that the input parameters are the cavity detuning $\Delta\Omega$ and the pump rates $Q_{A,B}$, which are both experimentally accessible.

We also solve SSCREs numerically and plot the solution versus $\Delta\Omega$, in Figure 1. Figs. 1(a) and 1(b) agree well with Equations (16) and (17), respectively, with Fig. 1(b) revealing detailed variations of $\Delta\omega$ beyond the first order approximate of Equation (17). Figs. 1(c) and 1(d) also agree well with Equations (11) and (14)-(15), respectively.

Tuning of the relative phase is expressed as $\sin\phi \cong \Delta\Omega/(2\alpha_H\kappa)$ in Equation (11). For each $\Delta\Omega$, there are two solutions of $\phi$, which are $\phi_+ = \arcsin\left(\frac{\Delta\Omega}{2\alpha_H\kappa}\right)$ and $\phi_- = \pi - \arcsin\left(\frac{\Delta\Omega}{2\alpha_H\kappa}\right)$. From the definition of the $\arcsin$ function, $\phi_+ \in [-\pi/2, \pi/2]$, while $\phi_- \in [\pi/2, 3\pi/2]$. When $\Delta\Omega = 0$, we have $\phi_+ = 0$ and $\phi_- = \pi$, as the in-phase and out-of-phase mode. When $\Delta\Omega \neq 0$, we have a tilted in-phase mode and tilted out-of-phase mode, labeled by + and – respectively. Other variables in the solution are labeled in accordance to $\phi$, making one solution set of $\left[\Delta\gamma_+, \Delta\omega_+, \left(\frac{Y_B}{Y_A}\right)_+, \phi_+\right]$ and the other set of $\left[\Delta\gamma_-, \Delta\omega_-, \left(\frac{Y_B}{Y_A}\right)_-, \phi_-\right]$.

The CRE analysis inherently has coupled mode theory embedded, so we can check consistency through the calculation of eigenmodes using couple mode theory with $\Delta\gamma_{+,-}$ and $\Delta\omega_{+,-}$ as input parameters. Coupled mode theory predicts two eigenmodes for $\Delta\gamma_+, \Delta\omega_+$ and another two for $\Delta\gamma_-, \Delta\omega_-$. However, only one out of the two eigenmodes for each set of $\Delta\gamma, \Delta\omega$ is consistent with the steady-state carrier rate equations, while the other eigenmode is not a stable solution. For example, if $\Delta\gamma_+, \Delta\omega_+$ are used as the input for coupled mode theory, the calculated eigenmodes are a tilted in-phase solution ($-\pi/2 < \phi < \pi/2$) and a tilted out-of-phase solution ($\pi/2 < \phi < 3\pi/2$). The tilted-in-phase solution satisfies Equation (4) automatically, while the tilted-out-of-phase solution does not. Similarly, for $\Delta\gamma_-, \Delta\omega_-$, only the tilted out-of-phase mode satisfies the carrier rate equation. Hence, for the optical mode to be a solution of the SSCREs, not only does the mode need to be a solution of coupled mode theory, it also needs to have a carrier density distribution that satisfies the rate equations.

When $|\Delta\Omega| > 2\alpha_H\kappa$, there are no steady-state solutions. Therefore, we can identity the cavity detuning range of $\Delta\Omega \in [-2\alpha_H\kappa, 2\alpha_H\kappa]$ to be the mutual injection locking range. From Equation (11) this can be understood as the requirement of $\sin\phi < 1$ for real $\phi$. To the best of our knowledge, this expression of the locking range first appeared in Ref. 21 and was later formally derived in Ref. 14.

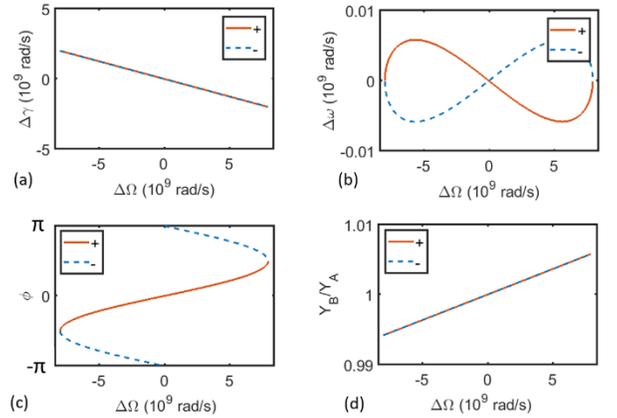

FIG. 1. Numerical solutions of the SSCREs for Array 1 (very weak coupling, $\kappa = 0.002/\tau_p$): (a) Induced gain contrast; (b) total frequency detuning; (c) relative phase; (d) field magnitude ratio between two cavities are plotted versus the cavity detuning $\Delta\Omega$. For $|\Delta\Omega| < 2\alpha_H\kappa$ there are two sets of solutions, labeled as + and – respectively. In (a) and (d) the two solutions are too close to distinguish in the plot. The pump parameters are set to $Q_A = Q_B = 3.2$, corresponding to $I_A = I_B = 2.375\ I_{th}$.

We consider the gain contrast induced by cavity detuning, shown in Fig. 1(a). This gain contrast consists of equal amount of optical gain and loss in the two cavities: $\gamma_A \cong \Delta\Omega/(2\alpha_H)$ and $\gamma_B = -\gamma_A$. The existence of loss is from the gain saturation. In other words, the optical loss arises from insufficient carrier density to maintain the excess amount of photons in the cavity. Intuitively, the connection between cavity detuning and the induced gain contrast can be understood as follows: with the existence of frequency detuning, the intensity distribution of the array eigenmodes becomes asymmetric, and this asymmetry in photon numbers in each cavity results in an asymmetric depletion rates of carriers (i.e. spatial hole burning). In turn, the carrier densities become asymmetric, which creates gain contrast. Mathematically, self-consistent solutions to the SSCREs are found to have equal gain and loss in each cavity while the frequency detuning is almost balanced out. We can also view this gain contrast between the lasers in terms of the energy flow and conservation of particle numbers. The lossy laser has greater photons emitted from its output mirror than the number of carrier injected, while the laser with net gain has fewer photons emitted than injected carriers. Therefore, there is net energy flow from the laser with gain into the lossy laser, and the laser with net gain supplies energy to the lossy laser through the coupled optical field, i.e. the array supermode. This energy transfer behavior does not exist in a Hermitian coupled array. It will be revisited in the next section, where we will see that the maximum magnitude of energy transfer scales with the coupling coefficient and it explains the different behavior in Array 2 compared to Array 1.

This cavity-detuning induced gain and loss suggests another way of reaching PT symmetry and exceptional points. In fact, in the limit of very weak coupling, the array under equal pumping nearly exhibits PT symmetry, in the sense that $\Delta\omega \cong 0$ to the first order of $\tau_p \kappa$. However, to reach exact PT symmetry and the exceptional points, tuning of the pump rates is necessary, as discussed in the following sections.

## IV. WEAKLY COUPLED ARRAYS UNDER UNEQUAL PUMPING

For unequal pumping into the two lasers, because a general analytical solution is not available, we solve SSCREs numerically using a numerical root search. The two cases of very weak coupling (Array 1) and moderate coupling (Array 2) are compared when the cavity detuning $\Delta\Omega$ and one of the pump rates $Q_B$ are varied. Evolution of the in-phase modes for Array 1 and 2 are plotted in Figures 2 and 3, respectively. The red lines show where the array is PT symmetric, which is discussed in greater detail in the next section.

In the case of very weak coupling, shown in Figure 2, from the color gradient we see that varying $Q_B$ has little effect on the gain contrast $\Delta\gamma$ or the relative phase $\phi$ (Figs. 2(a) and 2(c)), but it does control the total frequency detuning $\Delta\omega$ and the field magnitude ratio $(Y_B/Y_A)$ (Figs. 2(b) and 2(d)). The gain contrast and the relative phase are mostly controlled by the cavity detuning $\Delta\Omega$, evident from the color gradient in Figs. 2(a) and 2(c) being mostly along the horizontal direction. The solutions to the SSCREs for moderate coupling are shown in Figure 3. Similar to the case of very weak coupling, varying the pump parameter $Q_B$ still has little effect on gain contrast or phase tuning. However, the total frequency detuning $\Delta\omega$ and the field magnitude ratio are now controlled by both the $Q_B$ and $\Delta\Omega$, which is different from the case of very weak coupling.

For both Array 1 and Array 2, we find a finite region where steady-state solutions exist, which we identify as the locking region for the two lasers. Outside the locking region, no steady-state solution exists, which suggests either multi-mode lasing or temporally chaotic behavior [14, 35]. The horizontal width of the coupling region ($|\Delta\Omega|_{max}$) changes slightly with varying $Q_B$, but is approximately constant with $|\Delta\Omega|_{max} \cong 2\alpha_H \kappa$. This confirms that $\kappa \cong |\Delta\Omega_{max}|/2\alpha_H$ used in Ref. 21 is a good approximation for the experimental measurement of the coupling coefficient. The value $\Delta\Omega_{max}$ can be measured by adjusting the current into one of the lasers until the array emission breaks into multi-mode or instable operation.

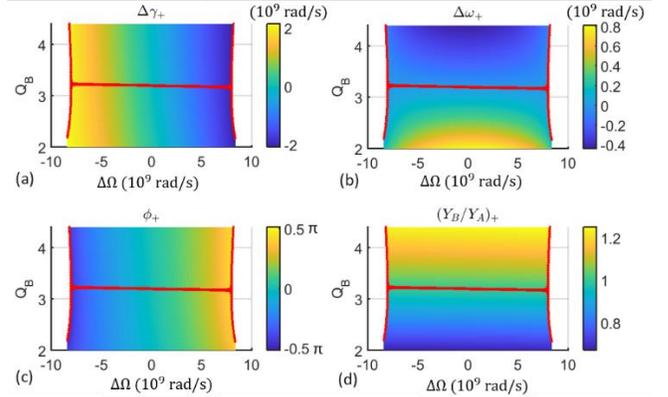

FIG. 2. Evolution of the tilted in-phase solution for Array 1 ($\kappa = 0.002/\tau_p$): (a) Induced gain contrast; (b) total frequency detuning; (c) relative phase; and (d) field magnitude ratio versus the cavity detuning and pump parameter $Q_B$, while $Q_A$ is fixed at 3.2. The pump parameters correspond to having $I_A$ fixed at 2.375 $I_{th}$, while $I_B$ varies from 1.625 $I_{th}$ to 3.125 $I_{th}$. Red lines show where the array is PT symmetric.

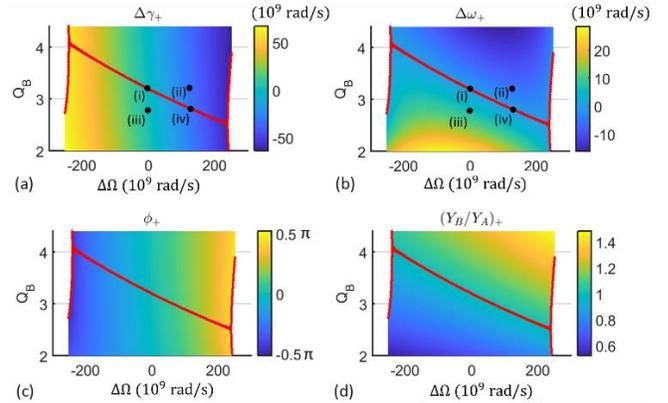

FIG. 3. Evolution of the tilted in-phase solution for Array 2 ($\kappa = 0.06/\tau_p$): (a) Induced gain contrast; (b) total frequency detuning; (c) relative phase; and (d) field magnitude ratio versus the cavity detuning and pump parameter $Q_B$. Again, $Q_A$ is fixed at 3.2, while $Q_B$ varies from 2 to 4.4. The locations labeled with numbers (i)-(iv) correspond to the schematics in Fig. 4. Red lines show where the array is PT symmetric.

The tilted out-of-phase solutions are plotted in Appendix A. They evolve similarly to the in-phase solutions plotted in Figures 2 and 3. With 2D plots like Figs. 2 and 3, we can determine the array mode evolution in response to the $\Delta\Omega$ and $Q_{A,B}$. In coupled VCSEL arrays, experimentally tuning the injected currents into each laser corresponds to varying both $Q_{A,B}$ and the $\Delta\Omega$ at the same time. The magnitude of injection currents not only change the pump parameters $Q_{A,B}$, but also varies the cavity resonance frequency $\Omega_{A,B}$ through ohmic heating and the refractive index temperature dependence. Hence varying the injection currents is equivalent to moving along a certain trajectory on the 2D maps shown in Figs. 2 and 3.

The different behaviors between very weak coupling (Array 1) and moderate coupling (Array 2) can be interpreted from the perspective of energy transfer. In the very weak coupling limit (Array 1 in Fig. 2), $Y_B/Y_A$ is almost solely determined by $Q_B/Q_A$, because the carrier injection rate (proportional to $Q_{B,A}$) needs to balance the carrier depletion rate, which is proportional to number of photons in the cavity (proportional to $Y_{B,A}^2$). However, when the optical coupling between cavities becomes stronger, the photon-mediated energy transfer between cavities can break this balance. For example, for Array 2 in Fig. 3, $Y_B^2/Y_A^2$ can be as large as 1.4 when $Q_B/Q_A = 1$, meaning that the photons in cavity B not only gains energy from the carriers injected into B, but also from the carriers injected into cavity A. This energy transfer is also connected to the gain/loss contrast between cavities. The cavity with more photons than injected carriers is interpreted as a lossy cavity, and it gains energy from the other cavity through optical coupling. The cavity that provides energy to the other cavity through optical coupling is interpreted to possess net gain.

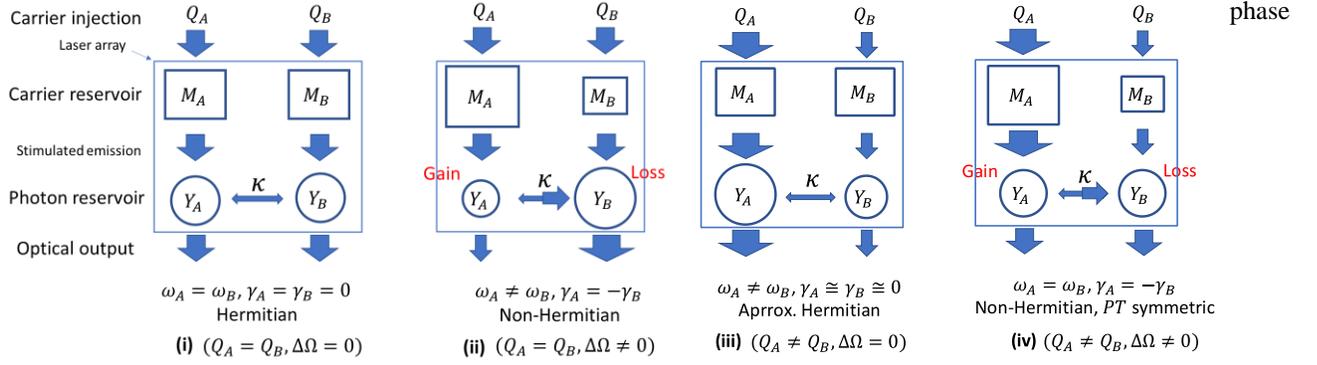

FIG. 4. Illustration of the distributions of carrier densities, photon densities, gain/loss, and energy flows at points labeled by (i)-(iv) in Fig. 3. Sizes of the boxes, circles, and arrows illustrate the asymmetry in carrier densities, photon densities, and energy flows (in the processes of carrier injection, stimulated emission, optical coupling, and optical output).

The various cases under equal and unequal pumping that can be considered are schematically shown in Figure 4. The sizes of arrows illustrate the magnitudes of energy flows associated with the processes of carrier injection, stimulated emission, and optical output from end mirrors of the cavities. The sizes of boxes and circles illustrate the carrier densities $M_{A,B}$ and photon densities $Y_{A,B}^2$. Four steady-state solutions (i)-(iv) are shown in Figure 4, which correspond to the four operating points labeled in Figure 3. Solution (i) is where the array is under equal pumping and zero cavity detuning. The array is Hermitian because there are no net gain or loss in either cavities. Solution (ii) is with equal pumping but nonzero cavity detuning $\Delta\Omega$, and the array is non-Hermitian due to gain/loss contrast induced by the cavity detuning. Although the pump rates into each cavities are the same in this situation, nonzero cavity detuning induces asymmetry in photon densities, which in turn affects the carrier depletion rate and results in asymmetric carrier densities. Also note the net energy flow from cavity A into cavity B through optical coupling. This energy flow is necessary for power conservation, which can be examined by summing up all the energy flows in and out of each reservoir. Solution (iii) is with nonequal pumping and zero cavity detuning. In this case, unequal pumping induces asymmetric carrier densities, which in turn induces frequency detuning through amplitude-coupling. The steady-state solution shows $\Delta\gamma \cong 0$, meaning that the array is approximately Hermitian. At last, solution (iv) is with judiciously chosen unequal pumping and cavity detuning that makes the array PT symmetric.

## V. PT SYMMETRY AND EXCEPTIONAL POINTS

For two identical resonators coupled through real coupling coefficient $\kappa$, the system is invariant under $\hat{P}\hat{T}$ if $\omega_A = \omega_B$, and $\gamma_A = -\gamma_B$ [22, 25, 30, 34, 36, 37]. However, the eigenmodes in the system may not be PT-invariant. It would be designated "unbroken PT symmetry" if both the system and the eigenmodes are PT-invariant. On the other hand, it would be designated "broken PT symmetry" when the system is PT-invariant but the eigenmodes are not. It is known that unbroken PT symmetry happens when $\Delta\gamma < 2\kappa$, while PT symmetry is spontaneously broken when $\Delta\gamma > 2\kappa$. At $\Delta\gamma = 2\kappa$, which is known as the exceptional points, the two eigenmodes collapse. Recently, improved sensing functionality has been demonstrated around the exceptional points [38, 39].

Points with $\omega_A = \omega_B$ are labeled in red in Figures 2 and 3, which correspond to where the array exhibits PT symmetry. In Figure 5, we specifically denote unbroken and broken PT symmetry regimes as blue and red lines; notice the exceptional points occur at their intersection. Here the gain contrast arises from equal gain and loss, meaning that it is naturally PT symmetric without the necessity of gauge transformation, for example in Ref. 40.

Along the line of unbroken PT symmetry, there are two sets of solutions to the SSCREs. At the exceptional points, the two sets of solutions collapse to the same values. Along the broken PT symmetry lines, there is only one set of solution to the SSCREs that satisfies both the coupled mode theory and the carrier density rate equations. Analytical solutions to the SSCREs are available along the line of unbroken PT symmetry, as discussed in Appendix B.

Operating the laser array at the exceptional point requires judiciously chosen pump ratio and cavity detuning. In most coupled diode laser arrays, since the pump ratio and cavity detuning are both controlled by the same experimental parameter, i.e. the injection currents, it could be challenging to find the exceptional point. However, the PT symmetry-breaking mode is relatively easy to achieve as long as there is sufficient cavity detuning to drive the array to the boundary of locking region. We note also that the broken PT symmetry can be achieved by exclusively cavity detuning (with equal pumping).

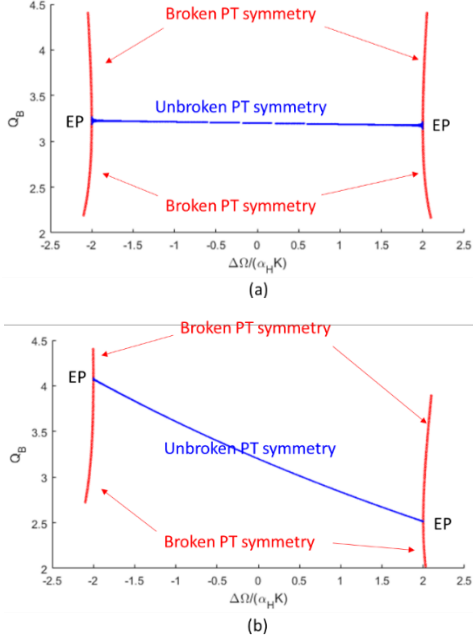

FIG. 5. Location of unbroken PT symmetry, broken PT symmetry and exceptional points (EPs) on the 2D parameter space of $(Q_B, \Delta\Omega)$ for (a) very weak coupling (Array 1) and (b) moderate coupling (Array 2).

A further observation is that the two sets of solutions to the SSCREs are generally different from each other, but they converge to the same solution along the lines of broken PT symmetry. Along the line of unbroken PT symmetry, the two sets of solutions share the same values of $\Delta\gamma, \Delta\omega, Y_B/Y_A$, but not $\phi$. Instead, they have $\phi_+ + \phi_- = \pi$. This observation is discussed more detail in Appendix A.

## VI. CONCLUSION

Mode tuning in coupled semiconductor lasers has been studied by solving the steady-state coupled rate equations. We show that depending on the strength of coupling compared to cavity loss rate, the coupled array responds differently to unequal pumping and cavity detuning. When $\kappa \ll 1/\tau_p$, which is the limit of very weak coupling, the cavity detuning induces a gain contrast, but the frequency detuning is almost completely balanced out by the frequency shift from the asymmetric carrier distribution. In the moderate coupling case ($\kappa = 0.06/\tau_p$), the frequency detuning is partially balanced out. In either weak or moderate coupling, gain contrast is more effectively introduced by the cavity detuning than by difference in pump rates, and the relative phase between two lasers are controlled by the cavity detuning, through the lever of induced gain contrast.

In the limit of very weak coupling, the tuning of intensity ratio between lasers is controlled almost solely by the pump rate difference, as expected from the conservation of energy and particle numbers in each cavity. In moderate coupling, because of the photon-mediated energy transfer between cavities, the particle number conservation should be considered in terms of the whole array instead of the individual cavities, and the intensity ratio is controlled by both the pump rate difference and $\Delta\Omega$.

To achieve unbroken PT symmetry or exceptional points in the semiconductor arrays with weak or moderate coupling, judiciously chosen cavity detuning and unequal pump rates are required. However, broken PT symmetry is less challenging to achieve, and it is possible to drive the array to PT symmetry breaking by exclusively cavity detuning.

The results presented have important implications for mode control in coupled semiconductor laser arrays, as well as the search for PT symmetry and exceptional points in such systems.

## ACKNOWLEDGEMENTS

Authors thank M J Adams and A Y Liu for helpful discussions. This research is supported by the National Science Foundation under Award No. ECCS 15-09845.

## APPENDIX A: OPTICAL MODE EVOLUTION OF COUPLED LASER ARRAYS

We discuss the evolution of the out-of-phase mode and the convergence of the tilted in-phase and tilted out-of-phase optical modes of coupled laser arrays. Similar to Figures 2 and 3 that show evolution of the in-phase mode, we plot evolution of the out-of-phase mode in Figure A1 and A2, for Array 1 and Array 2 respectively.

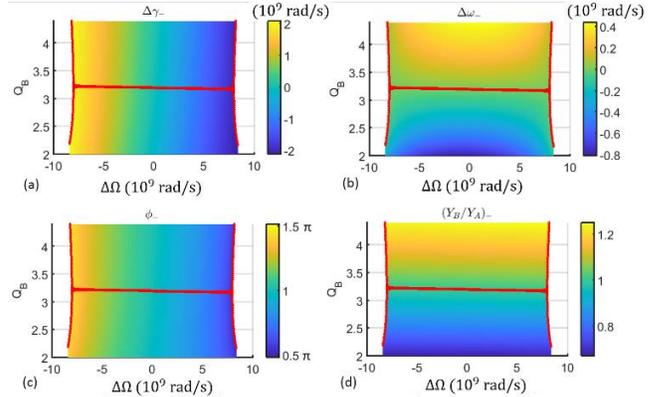

FIG. A1. Evolution of the tilted-out-of-phase solution for Array 1 ($\kappa = 0.002/\tau_p$): (a) Induced gain contrast; (b) total frequency detuning; (c) relative phase; and (d) field magnitude ratio versus the cavity detuning and pump parameter $Q_B$, while $Q_A$ is fixed at 3.2. The pump parameters correspond to having $I_A$ fixed at 2.375 $I_{th}$, while $I_B$ varies from 1.625 $I_{th}$ to 3.125 $I_{th}$. Red lines show where the array is PT symmetric.

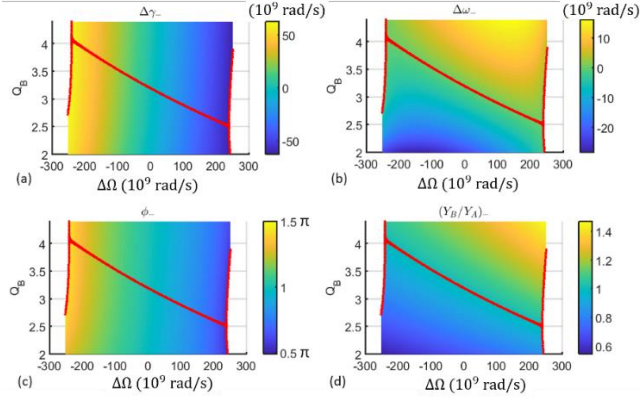

FIG. A2. Evolution of the tilted-out-of-phase solution for Array 2 ($\kappa = 0.06/\tau_p$): (a) Induced gain contrast; (b) total frequency detuning; (c) relative phase; and (d) field magnitude ratio versus the cavity detuning and pump parameter $Q_B$. Again, $Q_A$ is fixed at 3.2, while $Q_B$ varies from 2 to 4.4. Red lines show where the array is PT symmetric.

One further observation can be made by calculating the difference between the tilted out-of-phase mode and the tilted in-phase mode, namely $|\Delta\gamma_- - \Delta\gamma_+|$, $|\Delta\omega_- - \Delta\omega_+|$, $(\phi_- - \phi_+)$, and $\left|\left(\frac{Y_B}{Y_A}\right)_- - \left(\frac{Y_B}{Y_A}\right)_+\right|$, as shown in Figures A3 and A4. It can be observed that the two sets of solutions converge to the same value along the lines of broken PT symmetry, located at the boundary of the locking region. (For $(\phi_- - \phi_+)$, converges to $2\pi$ is equivalent to converging to 0.) Along the line of unbroken PT symmetry (see Figure 5), $\Delta\gamma, \Delta\omega,$ and $Y_B/Y_A$ from the two sets of solutions converge to the same value, but not $\phi$. From the property of the unbroken PT symmetric modes, we know that $\phi_+ + \phi_- = \pi$. In other words, $\Delta\gamma, \Delta\omega,$ and $Y_B/Y_A$ of the two sets of solutions converge when the array has either broken or unbroken PT symmetry, while $\phi_+$ and $\phi_-$ converge only when the array has broken PT symmetry. The mathematical structure of the solutions, which may be responsible for the converging behavior along the broken PT symmetry lines, is interesting for future study. It can be observed from the numerical solutions that $|\Delta\omega_+ - \Delta\omega_-|$ and $|\Delta\gamma_+ - \Delta\gamma_-|$ are linearly related and the line depicting broken PT symmetry might be a branch cut if we take linear combinations of $\Delta\omega$ and $\Delta\gamma$ to be the real and imaginary part of a complex variable.

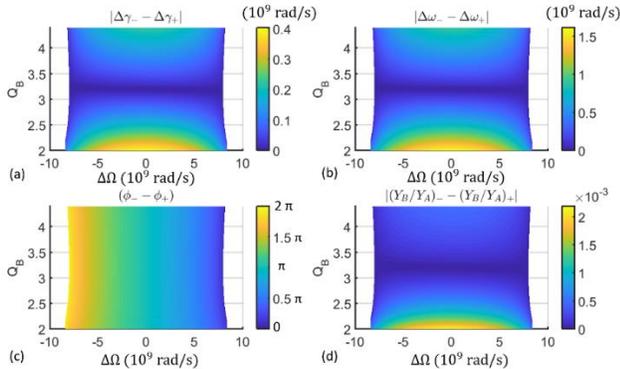

FIG. A3. Plot of the difference between the tilted-out-of-phase and tilted-in-phase solutions (Array 1, very weak coupling): (a) Absolute difference between the gain contrast $|\Delta\gamma_- - \Delta\gamma_+|$; (b) absolute difference between the total frequency detuning $|\Delta\omega_- - \Delta\omega_+|$; (c) difference between the relative phase $(\phi_- - \phi_+)$; (d) absolute difference between the field magnitude ratio $\left|\left(\frac{Y_B}{Y_A}\right)_- - \left(\frac{Y_B}{Y_A}\right)_+\right|$.

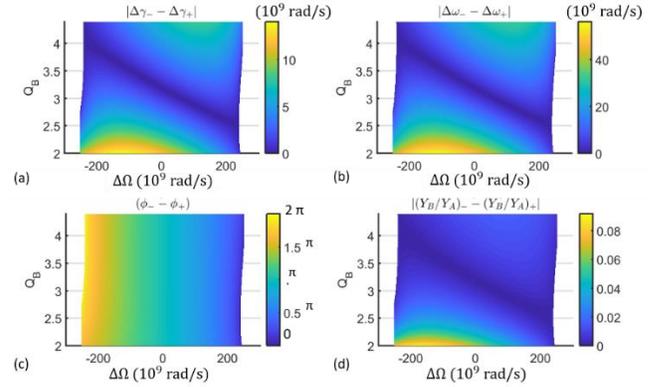

FIG. A4. Plot of the difference between the tilted-out-of-phase and tilted-in-phase solutions (Array 2, moderate coup): (a) Absolute difference between the gain contrast $|\Delta\gamma_- - \Delta\gamma_+|$; (b) absolute difference between the total frequency detuning $|\Delta\omega_- - \Delta\omega_+|$; (c) difference between the relative phase $(\phi_- - \phi_+)$; (d) absolute difference between the field magnitude ratio $\left|\left(\frac{Y_B}{Y_A}\right)_- - \left(\frac{Y_B}{Y_A}\right)_+\right|$.

Although anywhere along the lines of broken PT symmetry, we find the two solutions to SSCREs to collapse, this collapsing is different from the eigenmode collapsing at the exceptional points. At the exceptional points, the coupled mode equations predict two collapsed eigenmodes. While anywhere else along the lines of broken PT symmetry, the coupled mode equations predict two linearly independent eigenmodes, but only one of them satisfies the carrier rate equations.

## APPENDIX B: ANALYTICAL ANALYSIS UNDER UNBROKEN PARITY-TIME SYMMETRY

Condition of unbroken PT symmetry can be found analytically in the 2D parameter space of $Q_B$ and $\Delta\Omega$ (for example in Figures 2, 3, 5, A1 and A2):

$$\frac{Q_B}{Q_A} = \frac{1 - 2\tau_p\kappa\sin\phi}{1 + 2\tau_p\kappa\sin\phi}$$
$$\Delta\Omega = 2\alpha_H\kappa\sin\phi$$
$$\phi \in \left(-\frac{\pi}{2}, \frac{\pi}{2}\right)$$

Exceptional points are at the ends of the unbroken PT symmetry region, expressed as

$$\frac{Q_B}{Q_A} = \frac{1 \mp 2\tau_p\kappa}{1 \pm 2\tau_p\kappa}$$
$$\Delta\Omega = \pm 2\alpha_H\kappa$$

Along the line of unbroken PT symmetry, we have analytical solution to the steady-state coupled rate equations:

$$\Delta\omega_{+,-} = 0$$
$$\Delta\gamma_{+,-} = -\frac{\Delta\Omega}{\alpha_H}$$
$$\sin\phi = \frac{\Delta\Omega}{2\alpha_H\kappa}$$
$$M_A = 1 + 2\tau_p\kappa\sin\phi$$
$$M_B = 1 - 2\tau_p\kappa\sin\phi$$
$$Y_A^2 = Y_B^2 = \frac{1}{2}(Q_A + Q_B - 2)$$

This solution takes the same form as the approximate analytical solution for the weak-coupling equal-pumping array that was reported in Ref. 14 and repeated as (11)-(15) in Section III. This can be understood by noting that when the coupling coefficient goes to zero ($\tau_p \kappa \to 0$), the line of unbroken PT symmetry converges to the line of $Q_B = Q_A$.